\documentclass[twocolumn,showpacs,preprintnumbers,prd,superscriptaddress,nofootinbib]{revtex4-1}

\usepackage{amsmath}
\usepackage{amsfonts}
\usepackage{amssymb}
\usepackage{graphicx}
\usepackage{hyperref}
\usepackage{booktabs}
\usepackage{cleveref}
\usepackage{color}
\usepackage{mathrsfs} 
\usepackage{comment}

\Crefname{equation}{Eq.}{Eqs.}
\Crefname{figure}{Fig.}{Figs.}
\Crefname{section}{Sec.}{Secs.}
\Crefrangeformat{equation}{Eqs.~(#3#1#4)--(#5#2#6)}

\begin{document}

\title{Cosmographic view on the $H_0$ and $\sigma_8$ tensions}

\author{Rocco D'Agostino}
\email{rocco.dagostino@unina.it}
\affiliation{Scuola Superiore Meridionale (SSM), Largo S. Marcellino 10, I-80138 Napoli, Italy}
\affiliation{Istituto Nazionale di Fisica Nucleare (INFN), Sezione di Napoli, Via Cinthia 9, I-80126 Napoli, Italy}

\author{Rafael C. Nunes}
\email{rafadcnunes@gmail.com}
\affiliation{Instituto de F\'{i}sica, Universidade Federal do Rio Grande do Sul, 91501-970 Porto Alegre RS, Brazil}
\affiliation{Divis\~ao de Astrof\'isica, Instituto Nacional de Pesquisas Espaciais, Avenida dos Astronautas 1758, S\~ao Jos\'e dos Campos, 12227-010, SP, Brazil}

\begin{abstract}
Measurements of the $H_0$ and $\sigma_8$ parameters within the standard cosmological model recently highlighted significant statistical tensions between the cosmic microwave background and low-redshift probes, such as local distance ladder, weak lensing and galaxy clustering surveys. 
In this work, we frame geometrical distances in a model-independent way by means of cosmographic approximations in the range $z \in (0, 2.3)$ to take into account a robust dataset composed of Baryon Acoustic Oscillations (BAO),  type Ia Supernovae (SN),  Cosmic Chronometer (CC) data, and measurements from Redshift Space Distortions (RSD). From the joint analysis BAO\,+\,SN\,+\,CC\,+\,RSD, we find an accuracy of $\sim$1.4\% and $\sim$3.7\% on $H_0$ and $\sigma_8$, respectively. Our result for $H_0$ is at 2$\sigma$ tension with local measurements by the SH0ES team, while our $\sigma_8$ estimate is at 2.6$\sigma$ tension with Planck-CMB analysis. This inference shows a tension statistically smaller when compared to those estimated via the $\Lambda$CDM model. We also find that the jerk parameter can deviate more than 3$\sigma$ from the $\Lambda$CDM prediction. Under the same cosmographic setup, we also present results by considering a SH0ES gaussian prior on $H_0$ that allows for improved accuracy of the parameter space of the models. The present work brings observational constraints on $H_0$ and $\sigma_8$ into a new model-independent perspective, which differs from the predictions obtained within the $\Lambda$CDM paradigm.
\end{abstract}


\maketitle

\section{Introduction}
\label{sec:introduction}

Within the standard framework of general relativity, the acceleration of the universe observed today could be attributed to dark energy under the form of the cosmological constant ($\Lambda$), which drives the late-time cosmic evolution and whose origins are traced back to early quantum fluctuations of the vacuum \cite{SupernovaSearchTeam:1998fmf,SupernovaCosmologyProject:1998vns,Sahni:1999gb,Carroll:2000fy,Peebles:2002gy,Copeland:2006wr}. The flat $\Lambda$CDM model is supported by robust observational evidence promoting such a paradigm to be the standard model of cosmology \cite{WMAP:2012nax,Haridasu:2017lma,Planck:2018vyg}. 

However, despite its success, theoretical shortcomings related to the nature of $\Lambda$ on the one hand, and tensions among recent cosmological measurements, on the other hand, challenge the $\Lambda$CDM scenario as the ultimate model to describe the dynamics and evolution of the universe \cite{Zlatev:1998tr,Perivolaropoulos:2021jda}.
The fine-tuning issues emerging from the huge discrepancy between the observed dark energy density and the predictions of quantum field theory plague the standard interpretation of $\Lambda$ as the energy density of the vacuum \cite{Weinberg:1988cp,Zlatev:1998tr,Padmanabhan:2002ji,DAgostino:2019wko,DAgostino:2022fcx}.  Moreover, the latest findings of the Planck Collaboration have confirmed, up to a high accuracy level, that the most suitable scenario able to explain the primordial inflationary era is provided by the Starobinsky model \cite{Planck:2018jri}, which contemplates corrections with respect to Einstein's gravity.

On the other hand, some tensions and anomalies became recently significant when analyzing different datasets, placing the $\Lambda$CDM cosmology at a crossroads. The most discussed and statistically significant tension in the literature concerns the estimate of the Hubble constant, $H_0$, from the Cosmic Microwave Background (CMB) and the direct local distance ladder measurements. Assuming the $\Lambda$CDM scenario, Planck-CMB data analysis provides $H_0$ = (67.4 $\pm$ 0.5) km/s/Mpc \cite{Planck:2018vyg}, which is in $\sim$ 5$\sigma$ tension with the local measurement $H_0$ = (73.30 $\pm$ 1.04) km/s/Mpc found by the SH0ES team \cite{Riess:2021jrx}. Additionally, many other late-time measurements are in agreement with a higher Hubble constant value \cite{DiValentino:2021izs,DiValentino:2020zio}. Motivated by such discrepancies, unlikely to disappear completely by introducing multiple systematic errors, it has been widely discussed in the literature whether new physics beyond the standard cosmological model can solve the $H_0$ tension (see \cite{Adhikari:2022sve,Perivolaropoulos:2021jda,Schoneberg:2021qvd} and references therein). Still, in the context of the $\Lambda$CDM model, CMB measurements from Planck and ACT+WMAP indicate values of $S_8=0.834\pm0.016$ \cite{Planck:2018vyg} and $S_8=0.840\pm0.030$ \cite{ACT:2020gnv}, respectively, where $S_8\equiv\sigma_8\sqrt{\Omega_{m0}/0.3}$, being $\sigma_8$ the amplitude of matter fluctuations averaged on spheres of radius 8 Mpc/$h$, and $\Omega_{m0}$ the matter density parameter today. These values of $S_8$ are typically higher than those obtained by weak lensing and galaxy clustering measurements, which range between 0.703 to 0.782, showing a tension of $\sim$3$\sigma$ among these datasets\footnote{For a review and additional information on $S_8$ estimations, see \cite{DiValentino:2020vvd,Perivolaropoulos:2021jda,Adhikari:2022sve} and references therein.}. Although this tension might be due to systematic errors \cite{Amon_2022_s8}, it is worthwhile to investigate the possibility of new physics beyond the standard model to explain the $S_8$ tension \cite{Matteo_s8,Heisenberg:2022gqk,Poulin_s8,deAraujo:2021cnd}. Moreover, a large tension between Redshift Space Distortion (RSD) and CMB measurements has been identified by \cite{Skara:2019usd}. Disagreements between CMB and RSD measurements with other datasets, including the $E_G$ statistic, are discussed in detail by \cite{Nunes:2021ipq}, pointing out a tension up to 5$\sigma$.

The need to further investigate the nature of cosmic speed-up appears therefore essential in order to cure the aforementioned issues. At the same time, the degeneracy among different paradigms proposed over the last years to describe the dark energy behavior has motivated  the development of model-independent techniques that allow investigating the cosmic expansion without resorting to \emph{a priori} assumed cosmological setups \cite{Shafieloo:2005nd,Sahni:2008xx,Nesseris:2010ep,Daly:2003iy,LHuillier:2017ani}. 
Among these, of remarkable interest is the cosmographic approach \cite{Visser:2004bf,Cattoen:2008th,Dunsby:2015ers,Capozziello:2019cav,Yang:2019vgk,Lobo:2020hcz}, which relies only upon the cosmological principle and involves series expansions of the luminosity distance around the present time. Cosmography represents a powerful method as it provides a set of observable quantities that can be directly compared to data, and assures the independence of any postulated equation of state for dark energy. Hence, the cosmographic method has been widely used with the aim of breaking the degeneracy among different theoretical scenarios that behave in the same manner when compared to observations \cite{Capozziello:2008qc,Bamba:2012cp,Aviles:2012ay,Capozziello:2017uam,Capozziello:2022jbw,Rezaei:2020lfy}. 
Alternative robust methods to reconstruct the cosmological parameters in a model-independent way include, e.g., Gaussian process \cite{Holsclaw:2010sk,Shafieloo:2012ht,Seikel:2012uu,Gomez-Valent:2018hwc,Avila:2022xad,Perenon:2022fgw,Ruiz-Zapatero:2022zpx,OColgain:2021pyh}, principal component analysis \cite{Huterer:2002hy,Crittenden:2005wj,Zhao:2017cud,Sharma:2020unh} or machine learning algorithms \cite{Fluri:2019qtp,Arjona:2019fwb,Escamilla-Rivera:2019hqt}.

Nevertheless, the cosmographic technique is affected by two main problems that may limit its use as an accurate descriptor of cosmic expansion. The first is due to the need for a wide number of data in order to be able to properly distinguish between $\Lambda$ and an evolving dark energy component. This, indeed, is required for reducing the uncertainties over the cosmographic coefficients. A second issue is related to the use of high-redshift data, which is needed to explore possible departures from the $\Lambda$CDM model. However, this contrasts with the foundation of the standard cosmographic technique, which is based on the Taylor expansion series around the present time, namely $z=0$. The resulting convergence problems often lead to significant error propagations that, consequently, lower the predictive power of the method itself \cite{Busti:2015xqa}.

Over the years, several alternatives to the standard cosmographic approach have been then investigated with the purpose of overcoming the aforementioned drawbacks. One possibility is to make use of auxiliary variables and provide expansion series of cosmological observables in terms of re-parametrizations of the redshift that converge to a finite value for $z\rightarrow \infty$ \cite{Cattoen:2007sk,Aviles:2012ay,Risaliti:2018reu,Bargiacchi:2021fow}. 
Further methods involve the use of rational polynomials to stabilize the behavior of the cosmographic series at large $z$ \cite{Capozziello:2017nbu}. A relevant example of the latter class is offered by the Pad\'e approximations, which have been recently employed to study the nature of the cosmic acceleration in different theoretical contexts, due to their ability to overcome typical convergence issues proper of the Taylor series and significantly reduce uncertainties on fitting coefficients \cite{Wei:2013jya,Gruber:2013wua,Capozziello:2017ddd,Capozziello:2018aba}.

The aim of the present work is to assess cosmological tensions in the measurements of $H_0$ 
and $\sigma_8$ from a model-independent perspective through cosmography. In particular, we adopt the $y$-redshift parametrization proposed by \cite{Cattoen:2007sk} and the (2,1) Pad\'e approximation, motivated by previous results obtained by \cite{Capozziello:2020ctn}. We describe the main features of the cosmographic technique based on these parametrizations in Section \ref{sec:model}. In Section \ref{sec:data}, we present the methodology and the datasets employed to analyze the $H_0$ and $\sigma_8$ tensions. Then, in Section \ref{sec:results}, we provide the main results of this work and discuss possible implications in view of the state of the art of the aforementioned tensions. Finally, Section \ref{sec:conclusions} is dedicated to the summary of our findings and conclusive remarks. In this paper, we use natural units, $c=\hbar=1$.

\section{The cosmographic approach}
\label{sec:model}

The global evolution of the universe can be studied by requiring the validity of the cosmological principle, according to which the universe is assumed to be isotropic and homogeneous on the largest scales.
Such a principle is supported by an overwhelming number of observations and formally leads to the Friedmann-Robertson-Walker metric:
\begin{equation}
ds^2= dt^2 -a(t)^2 \left[dr^2+r^2(d\theta^2+\sin^2\theta\, d\phi^2)\right],
\end{equation}
where a vanishing spatial curvature is assumed, as suggested by observations \cite{Planck:2018vyg}.
Here, $a(t)$ is the cosmic scale factor, normalized such that $a(t_0)=1$, being $t_0$ the present time.

Differently from standard cosmological approaches based on the solutions of the Friedmann equations, cosmography  allows for a kinematic study of the cosmic expansion that is totally independent of the underlying dynamics governing the universe's evolution. Thus, by means of the cosmographic method, one can infer the history of $a(t)$ directly from observations, \emph{de facto} avoiding the use of Einstein's field equations \cite{Visser:2004bf}. 

The cosmographic method is based on the Taylor expansion of the scale factor around the present time \cite{Weinberg:1972kfs}:
\begin{align}
    a(t)=&\ 1+H_0(t-t_0)-\frac{1}{2}q_0 H_0^2(t-t_0)^2+\frac{1}{3!} j_0 H_0^3 (t-t_0)^3 \nonumber  \\
    &+\frac{1}{4!}s_0 H_0^4(t-t_0)^4+\frac{1}{5!}l_0 H_0^5(t-t_0)^5 + \mathcal{O}\left((t-t_0)^6\right),
\end{align}
where $H_0$ is the Hubble constant and $\{q_0,j_0,s_0,l_0\}$ are the current values of the deceleration, jerk, snap and lerk parameters, respectively, defining the so-called cosmographic series \cite{Visser:2003vq}:
\begin{align}
&H(t)\equiv \dfrac{1}{a}\dfrac{da}{dt}\,, \quad q(t)\equiv -\dfrac{1}{aH^2}\dfrac{d^2a}{dt^2}\,, \quad  j(t) \equiv \dfrac{1}{aH^3}\dfrac{d^3a}{dt^3}\,, \nonumber  \\  
&s(t)\equiv\dfrac{1}{aH^4}\dfrac{d^4a}{dt^4}\ , \quad    l(t)\equiv \frac{1}{a H^5} \frac{d^5a}{dt^5}\,.
\label{eq:CS}
\end{align}
The ability to discriminate among different dark energy models should, in principle, increase when considering higher-order terms. However, the lack of large and accurate observational data at high redshifts somewhat restricts the use of cosmographic coefficients up to the lerk parameter.

Moving to the redshift variable through the relation $a=(1+z)^{-1}$, one can write the luminosity distance for a flat universe as
\begin{equation}
    d_L(z)=(1+z)\int_0^z \frac{dz'}{H(z')}\,.
    \label{eq:dL}
\end{equation}
In view of the above definitions, the latter provides
\begin{align}
d_L(z)=&\ H_0^{-1}\bigg[z +  \dfrac{1}{2}(1 - q_0) z^2 - \dfrac{1}{6}(1 - q_0 - 3 q_0^2 + j_0) z^3 \nonumber \\
&+ \dfrac{1}{24}(2 - 2 q_0 - 15 q_0^2 - 15 q_0^3 + 5 j_0 +10 q_0 j_0 +s_0) z^4 \nonumber \\
&+  \left(-\dfrac{1}{20} - \dfrac{9j_0 }{40}+ \dfrac{j_0^2}{12} - \dfrac{l_0}{120} + \dfrac{q_0}{20} - \dfrac{11 j_0 q_0}{12} + \dfrac{27 q_0^2}{40}\right.  \nonumber  \\
& \left. - \dfrac{7 j_0 q_0^2}{8} + \dfrac{11 q_0^3}{8} + \dfrac{7 q_0^4}{8} - \dfrac{11 s_0}{120} -\dfrac{q_0 s_0}{8}\right) z^5\bigg]+\mathcal{O}(z^6)\,,
\label{eq:dL Taylor}
\end{align}
which can be used to describe the late-time evolutionary stage of the universe without any assumptions on the cosmological model. Furthermore, by inverting Eq.~\eqref{eq:dL}, it is possible to find the corresponding Hubble series.

Comparing Eq.~\eqref{eq:dL Taylor} directly with observations provides numerical bounds over the cosmographic parameters and, thus, allows one to reconstruct the cosmic expansion history up to a desired $z$. 
However, it is worth to remark that truncating the cosmographic series at a given order may cause biases in the numerical outcomes. If, on one hand, taking into account only low-order expansions may compromise the accuracy of the method, on the other hand, considering higher orders induces decreasing convergence. The latter issue occurs when data at $z> 1$ are employed in the analysis, and is a consequence of the short convergence radius of Taylor series. In what follows, we shall face this problem by considering two different strategies aimed at extending the suitability of the cosmographic technique up to high redshifts.

\subsection{$y$-redshift}

The first possibility to overcome the convergence problem is to introduce auxiliary variables and re-parametrize the redshift via a function that well behaves for $z>1$. These new variables must possess some suitable properties, such as exhibiting smooth features throughout cosmic history thus avoiding any divergence within the domain, and being one-to-one invertible with the original $z$ variable.

A relevant example in this respect is provided by the so-called $y$-redshift introduced by \cite{Cattoen:2007sk}:
\begin{equation}
    y=\frac{z}{1+z}\,,
\end{equation}
which allows encoding the cosmic evolution back to the Big Bang into the finite range $y\in(0,1)$. Therefore, we expect that the luminosity distance expressed as a Taylor series in powers of $y$ well behaves from the present time up to early epochs, and the convergence radius to be $|y|=1$, implying the convergence of the series for $|y|<1$.

In terms of the variable $y$, up to the fifth order, we have 
\begin{align}
d_L(y)=&\ H_0^{-1}\bigg[y + \dfrac{1}{2} (3- q_0) y^2 + \dfrac{1}{6} (11 - j_0 - 5 q_0 + 3 q_0^2) y^3 \nonumber  \\
&+ \dfrac{1}{24} (50 - 7 j_0 - 26 q_0 + 10 j_0 q_0  + 21 q_0^2 - 15 q_0^3 + s_0) y^4 \nonumber \\
&+ \dfrac{1}{120} (274 - 47 j_0 + 10 j_0^2 - l_0 - 154 q_0 + 90 j_0 q_0 + 141 q_0^2 \nonumber \\
&- 105 j_0 q_0^2 - 135 q_0^3 + 105 q_0^4 + 9 s_0 - 15 q_0 s_0) y^5\bigg],
\end{align}
while the Hubble parameter is given as
\begin{eqnarray}
\label{H_y}
H(y)=H_0\left(1+k_1 y+\dfrac{k_2y^2}{2}+\dfrac{k_3y^3}{6}+\dfrac{k_4y^4}{24}\right),
\end{eqnarray}
where
\begin{subequations}
\begin{align}
k_1&=1+q_0\,, \\
k_2&=2-q^2_0+2q_0+j_0\,, \\
k_3&=6+3q^3_0-3q^2_0+6q_0-4q_0j_0+3j_0-s_0\,, \\
k_4&=-15q^4_0+12q^3_0+25q^2_0j_0+7q_0s_0-4j^2_0-16q_0j_0-12q^2_0\nonumber \\
&\hspace{0.4cm}+l_0-4s_0+12j_0+24q_0+24\,.
\end{align}
\end{subequations}

\subsection{Pad\'e parametrizations}

The issue of stabilizing cosmographic expansions at high $z$ can be also addressed by making use of rational approximations. These are constructed from the ratio between a generic $n$-th degree polynomial and a $m$-th degree polynomial, leading to $(n,m)$-order polynomials that can be used to approximate cosmic observables in terms of the cosmographic coefficients. The advantage of rational approximations relies on the fact that they can be calibrated by choosing the most suitable order for maximizing the convergence radius, thus allowing for a stable fitting procedure.

The good properties of rational polynomials are testified by the case of Pad\'e approximations, which have been shown to overcome the high-redshift divergences plaguing the cosmographic analysis based on Talyor polynomials \cite{Aviles:2014rma}. The $(n,m)$ Pad\'e approximation of a generic function $f(z)$ is given as \cite{baker1996pade}
\begin{equation}
P_{n, m}(z)=\dfrac{\displaystyle{\sum_{i=0}^{n}a_{i} z^{i}}}{1+\displaystyle{\sum_{j=1}^{m}b_j z^{j}}}\,,
\end{equation}
where the sets of coefficients $a_i$ and $b_i$ can be found by requiring the matching of the $n+m$ derivatives of $P_{n,m}$ evaluated at the origin and the corresponding derivatives obtained from the Taylor series of $f(z)$. 

The matter of choosing the correct order of expansion is related to the best compromise between minimizing the number of free parameters and reducing error propagation in numerical analyses dealing with data beyond $z\simeq 1$. This issue has been recently addressed by \cite{Capozziello:2020ctn}, who performed a detailed study based on optimization procedures and mathematical considerations on the degeneracy among coefficients,  showing that the most suitable Pad\'e approximation for cosmographic purposes is the one of order (2,1). The latter is characterized by the following luminosity distance:
\begin{equation}
d_L(z)=\dfrac{1}{H_0}\left[\dfrac{6 (-1 + q_0) z+(-5 - 2 j_0 + q_0 (8 + 3 q_0)) z^2}{6(-1 + q_0)+2(-1 - j_0 + q_0 + 3 q_0^2) z}\right].
\end{equation}
In this case, the Hubble parameter is given by
\begin{equation}
\label{H_pade}
    H(z)=\dfrac{2 H_0 (z+1)^2 \left(j_0 z-3 q_0^2 z-q_0 (z+3)+z+3\right)^2}{ p_0+p_1 z+p_2 z^2}\,,
\end{equation}
where 
\begin{subequations}
\begin{align}
    & p_0=18 (-1 + q_0)^2\,, \\
    & p_1=6 (-1 + q_0) (-5 - 2 j_0 + 8 q_0 + 3 q_0^2)\,, \\
    & p_2=14 + 7 j_0 + 2 j_0^2 - 10 (4 + j_0) q_0 + (17 - 9 j_0) q_0^2 + 18 q_0^3 + 9 q_0^4\,.
\end{align}
\end{subequations}

Motivated by the aforementioned arguments, in section \ref{sec:results} we shall take into account the $y$-redshift and (2,1) Pad\'e approximations to investigate the expansion history of the Universe in a model-independent way so as to gain further insights into the cosmological tensions.

\section{Datasets and Methodology}
\label{sec:data}

In the following, we define the datasets that will be used in our analysis. 

\begin{itemize}

\item \textbf{BAO}. From the latest compilation of Baryon Acoustic Oscillations (BAO) distance and expansion rate measurements from the SDSS collaboration, we use 14 BAO measurements, viz., the isotropic BAO measurements of $D_V(z)/r_d$ (where $D_V(z)$ and $r_d$ are the spherically averaged volume distance, and sound horizon at baryon drag, respectively) and anisotropic BAO measurements of $D_M(z)/r_d$ and $D_H(z)/r_d$ (where $D_M(z)$ and $D_H(z)=c/H(z)$ are the comoving angular diameter distance and  the Hubble distance, respectively), as compiled in Table 3 of \cite{eBOSS:2020yzd}. This measurement sample covers the range $z \in$ [0.15, 2.33], where the line-of-sight comoving distance is calculated by assuming the expansion rate of the universe given from the cosmographic expressions (\ref{H_y}) and (\ref{H_pade}).

\item \textbf{SN}. We also consider the type Ia Supernovae (SN) distance moduli measurements from the \textit{Pantheon} sample, consisting of 1048 SNeIa in the range $0.01<z<2.3$~\cite{Scolnic:2017caz}, used to constrain the normalized expansion rate $E(z) = H(z)/H_0$ \cite{Riess:2017lxs}.   

\item  \textbf{CC}. Our analysis involves the cosmic chronometer measurements of $H(z)$ from the differential age evolution of massive, early-time, passively evolving galaxies acting as standard clocks~\cite{Jimenez:2001gg}. In particular, we make use of 31 CC measurements of $H(z)$ in the range $0.07<z<1.965$, compiled by \cite{Jimenez:2003iv,Simon:2004tf,Stern:2009ep,Moresco:2012by,Zhang:2012mp,Moresco:2015cya,Moresco:2016mzx,Ratsimbazafy:2017vga}.

\item \textbf{SH0ES}. A gaussian prior on the Hubble constant as measured by the SH0ES collaboration \cite{Riess:2021jrx} is taken into account, i.e., $H_0 = (73.04 \pm 1.04)$ km/s/Mpc.

\item \textbf{RSD}. The key dataset we use to constrain the parameter space $\sigma_8$-$H_0$ through cosmography is the redshift space distortion measurements. These represent a velocity-induced mapping from the real to the redshift space due to line-of-sight peculiar motions of objects, which introduce anisotropies in their clustering patterns~\cite{Kaiser:1987qv}. This effect depends on the growth of structure, making RSD probes sensitive to the following combination \cite{Nesseris:2017vor,DAgostino:2018ngy}:
\begin{equation}
     f\sigma_8(a)\equiv f(a)\sigma_8(a)\,,
\end{equation}
where $f(a)\equiv d\delta_m/d\ln a$, with $\delta_m$ being the linear matter density contrast. Here, $\sigma_8(a)\equiv\sigma_{8} \delta_m(a)/\delta_m(1)$ is the linear amplitude of matter fluctuations averaged in spheres of radius 8 $h^{-1}$Mpc, and $\sigma_{8}$ its present-day value.
On sub-horizon scales and in the linear regime, the evolution equation for $f(a)$ is given by
\begin{equation}
\frac{df(a)}{d\ln a} + f(a)^2 + \left ( 2 + \frac{1}{2} \frac{d \ln H(a)^2}{d \ln a} \right ) f(a)- \frac{3}{2} \Omega_m(a) = 0\,,
\end{equation}
where $\Omega_m(a) \equiv \Omega_{m0} a^{-3}H_0^2/H(a)^2$.
Notice that $f(a)$ is dependent on $H(a)$ and, thus, on the cosmographic parameters. Since $\Omega_{m0}=0.31$ is the mean value used in obtaining almost all RSD measurements, without loss of generality, we assume this value in our work, when RSD data are used. Due to the cosmographic approximations, all the other datasets have no dependence on $\Omega_{m0}$, so it will not be possible to break the degeneracy in the matter density when they are combined with RSD data. We checked out that by letting $\Omega_{m0}$ free, this parameter becomes unconstrained without affecting the best-fit values of the cosmographic baseline parameters.

Several measurements of $f\sigma_8(a)$ from a variety of different surveys, based on different assumptions (in particular, on the reference value of $\Omega_{m0}$) and subject to different systematics, exist in the literature. Before using any of them, it is imperative to assess their internal consistency. Such an analysis has been recently performed by \cite{Sagredo:2018ahx} in the context of a Bayesian model comparison framework. There, it was possible to identify potential outliers as well as subsets of data affected by systematics or new physics. In this work, we shall make use of the RSD measurements of $f\sigma_8(z)$ provided in Tab.~I by \cite{Sagredo:2018ahx}, consisting of 22 measurements of $f\sigma_8(z)$ in the redshift range $0.02<z<1.944$.

\end{itemize}

\subsection{MCMC}
\label{sec:methodology}

We use the Markov Chain Monte Carlo (MCMC) method to analyze the parameter set $\theta_i = \{H_0, q_0, j_0, l_0, s_0, \sigma_8 \}$, building  the posterior probability distribution function 
\begin{equation}
\label{L}
\mathcal{P}(D|\theta) \propto e^{- \frac{\chi^2}{2}}\,,
\end{equation}
where $\chi^2$ is the chi-square function associated with each dataset.
The goal of any MCMC approach is to draw $N$ samples $\theta_i$ from the general posterior probability density
\begin{equation}
\label{psd}
\mathcal{P}(\theta_i, \alpha|D)=\frac{1}{Z}\mathcal{P}(\theta,\alpha)\mathcal{P}(D|\theta,\alpha) \,,
\end{equation}
where $\mathcal{P}(\theta,\alpha)$ and $\mathcal{P}(D|\theta,\alpha)$ are the prior distribution and the likelihood function, respectively. Here, $D$ refers to the dataset, $\alpha$ accounts for possible nuisance parameters, and $Z$ is a normalization term.

We perform our statistical analysis by means of the \texttt{emcee} algorithm \cite{emcee}, assuming the theoretical setups described in Sec.~\ref{sec:model} and the following flat priors: $H_0 \in [10, 90]$, $q_0 \in [-2, 0]$, $j_0 \in [-10, 10]$, $s_0 \in [-100, 100]$, $l_0 \in [-100, 100]$ and $\sigma_8 \in [0.5, 1.5]$. We discard the first 20\% steps of the chain as burn-in. We measure the convergence of the chains by checking that all parameters have $R - 1 < 0.01$, where $R$ is the potential scale reduction factor of the Gelman-Rubin diagnostics \cite{Gelman:1992zz}. The output from the chains is analyzed through the package \texttt{ChainConsumer} \cite{Hinton2016}.

Under the cosmographic approach, each analysis involves at least 3 free parameters. Thus, given the dimension of the parameter space, we consider BAO\,+\,SN\,+\,CC as our minimal baseline. We divided our analysis into two steps:
\begin{itemize}

\item First, we analyze the BAO\,+\,SN\,+\,CC case. Then, we include SH0ES prior, i.e., BAO\,+\,SN\,+\,CC\,+\,SH0ES. 

\item Second, we add the RSD measurements to the minimal baseline, i.e.,  BAO\,+\,SN\,+\,CC\,+\,RSD. Then, we also analyze the BAO\,+\,SN\,+\,CC\,+\,RSD\,+\,SH0ES combination.

\end{itemize}

The above joint analyses provide us with an overview of the observational constraints on the free parameters of both models under study in this work. In what follows, we present our main results.

\begin{table*}
    \centering
    \caption{68\% C.L. intervals on the cosmographic parameters within the $y$-redshift parametrization inferred from different combinations of datasets. The $H_0$ values are given in units of km/s/Mpc.}
    \label{tab:mmodel1}
    \setlength{\tabcolsep}{0.7em}
    \renewcommand{\arraystretch}{2}
    \begin{tabular}{c|cccccc}
        \hline
		Datasets & $H_0 $ & $q_0$ & $j_0$ & $s_0$ & $l_0$ & $\sigma_8$ \\ 
		\hline
		BAO+SN+CC & $69.21^{+0.97}_{-1.05}$ & $-0.56^{+0.20}_{-0.17}$ & $-0.4^{+2.0}_{-2.5}$ & $-12^{+11}_{-18}$ & $>-30$ & -- \\ 
		BAO+SN+CC+SH0ES & $71.09^{+0.74}_{-0.77}$ & $-0.69^{+0.16}_{-0.18}$ & $0.4\pm 2.0$ & $-8^{+10}_{-15}$ & $>-30$ & -- \\ 
		BAO+SN+CC+RSD & $69.21^{+0.91}_{-1.10}$ & $-0.54^{+0.17}_{-0.20}$ & $-0.2^{+1.8}_{-2.3}$ & $-7.7^{+6.3}_{-19.7}$ & $>-28$ & $0.737^{+0.029}_{-0.027}$ \\ 
		BAO+SN+CC+RSD+SH0ES & $71.02^{+0.88}_{-0.67}$ & $-0.70^{+0.12}_{-0.21}$ & $0.8\pm 1.8$ & $-1.9^{+3.1}_{-18.4}$ & $>-30$ & $0.725^{+0.026}_{-0.029}$ \\ 
		\hline
    \end{tabular}
\end{table*}

\begin{table*}
    \centering
    \caption{68\% C.L. intervals on the cosmographic parameters within the (2,1) Pad\'e approximation inferred from different combinations of datasets. The $H_0$ values are given in units of km/s/Mpc.}
    \label{tab:mmodel2}
    \setlength{\tabcolsep}{0.7em}
    \renewcommand{\arraystretch}{2}
    \begin{tabular}{c|cccc}
        \hline
		Datasets & $H_0$ & $q_0$ & $j_0$ & $\sigma_8$ \\ 
		\hline
        BAO+CC & $66.5^{+2.0}_{-1.9}$ & $-0.38^{+0.16}_{-0.21}$ & $0.59^{+0.93}_{-0.53}$ & -- \\ 
		BAO+SN+CC & $69.11^{+1.06}_{-0.98}$ & $-0.663^{+0.088}_{-0.095}$ & $2.06^{+0.60}_{-0.51}$ & -- \\ 
		BAO+SN+CC+SH0ES & $71.06^{+0.81}_{-0.71}$ & $-0.803^{+0.079}_{-0.084}$ & $2.88^{+0.63}_{-0.56}$ & -- \\ 
		BAO+SN+CC+RSD & $69.2\pm 1.0$ & $-0.663^{+0.086}_{-0.094}$ & $2.05^{+0.59}_{-0.49}$ & $0.740^{+0.027}_{-0.029}$ \\ 
		BAO+SN+CC+RSD+SH0ES & $71.09^{+0.73}_{-0.75}$ & $-0.797^{+0.077}_{-0.084}$ & $2.84^{+0.61}_{-0.54}$ & $0.724^{+0.028}_{-0.026}$ \\ 
		\hline
    \end{tabular}
\end{table*}

\section{Results and Discussion}
\label{sec:results}

Let us start by discussing the results emerging from the perspective of the $y$-redshift cosmography. 
We summarize in Table \ref{tab:mmodel1} our results at the 68\% confidence level (C.L.). We note that the first parameters of the cosmographic series, namely $H_0$, $q_0$ and $j_0$, are well constrained. The model-independent estimate of the Hubble constant is $H_0=69.21^{+0.97}_{-1.05}$ km/s/Mpc, with 1.4\% accuracy. This result is competitive with other current estimates  performed by assuming the $\Lambda$CDM cosmology
 \cite{Schoneberg:2022ggi,Philcox:2020vvt,Chen:2021wdi,Nunes:2020hzy}. Also, the deceleration parameter is robustly constrained to $q_0=-0.56^{+0.20}_{-0.17}$, suggesting the current accelerated expansion of the universe at high statistical significance evidence. When the SH0ES prior is taken into account, the constraints over the Hubble constant are further improved: $H_0=71.09^{+0.74}_{-0.77}$ km/s/Mpc, with 1\% accuracy.

\begin{figure*}
\begin{center}
\includegraphics[width=7.8in]{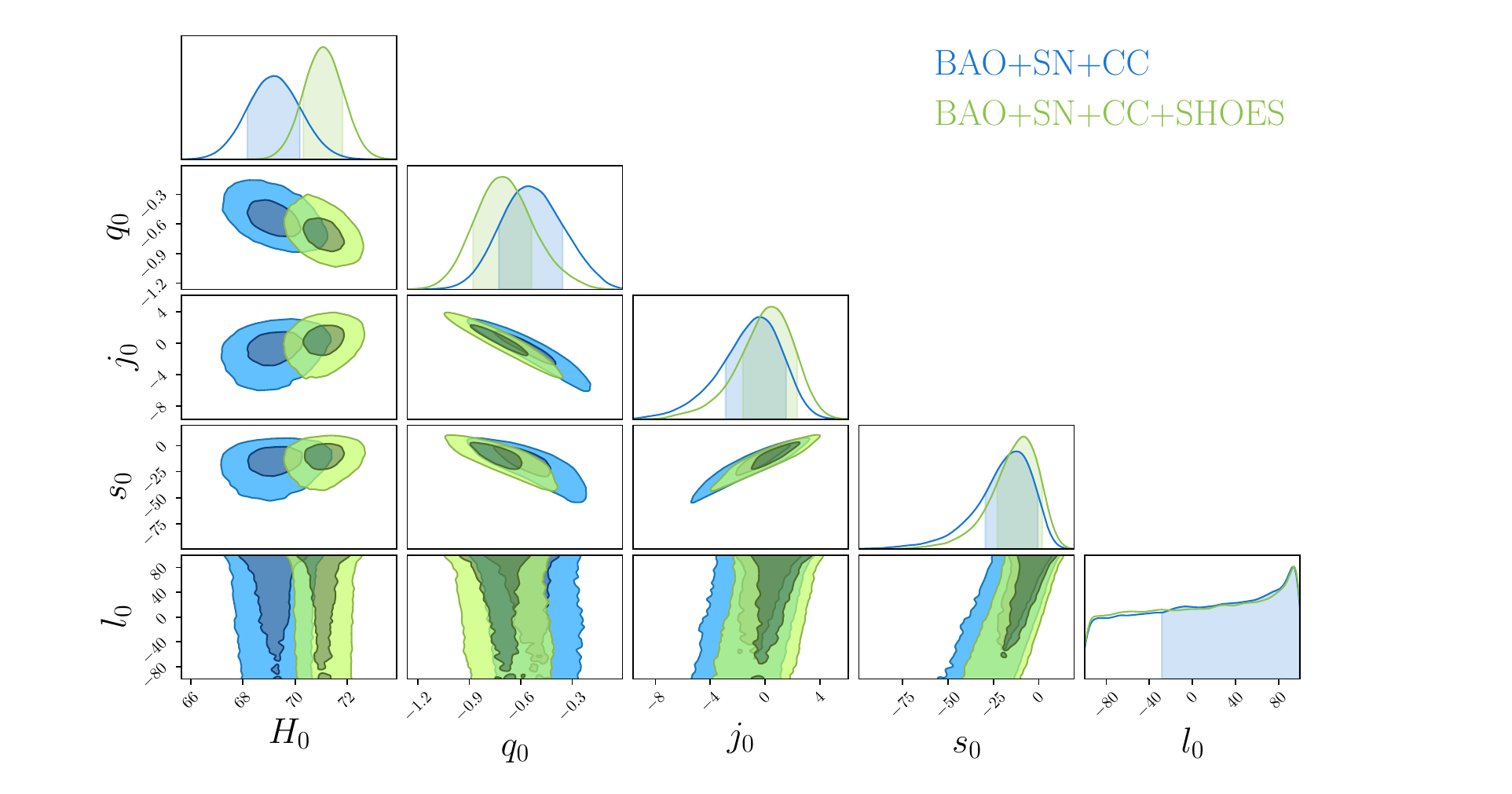} 
\caption{Two-dimensional marginalized confidence regions (68\% and 95\% C.L.) and one-dimensional posterior distribution for the cosmographic coefficients obtained from the joint BAO\,+\,SN\,+\,CC and BAO\,+\,SN\,+\,CC\,+\,SH0ES analyses for the $y$-redshift parametrization. The $H_0$ values are expressed in units of km/s/Mpc.}
\label{fig:base1}
\end{center}
\end{figure*}

\begin{figure*}
\begin{center}
\includegraphics[width=7in]{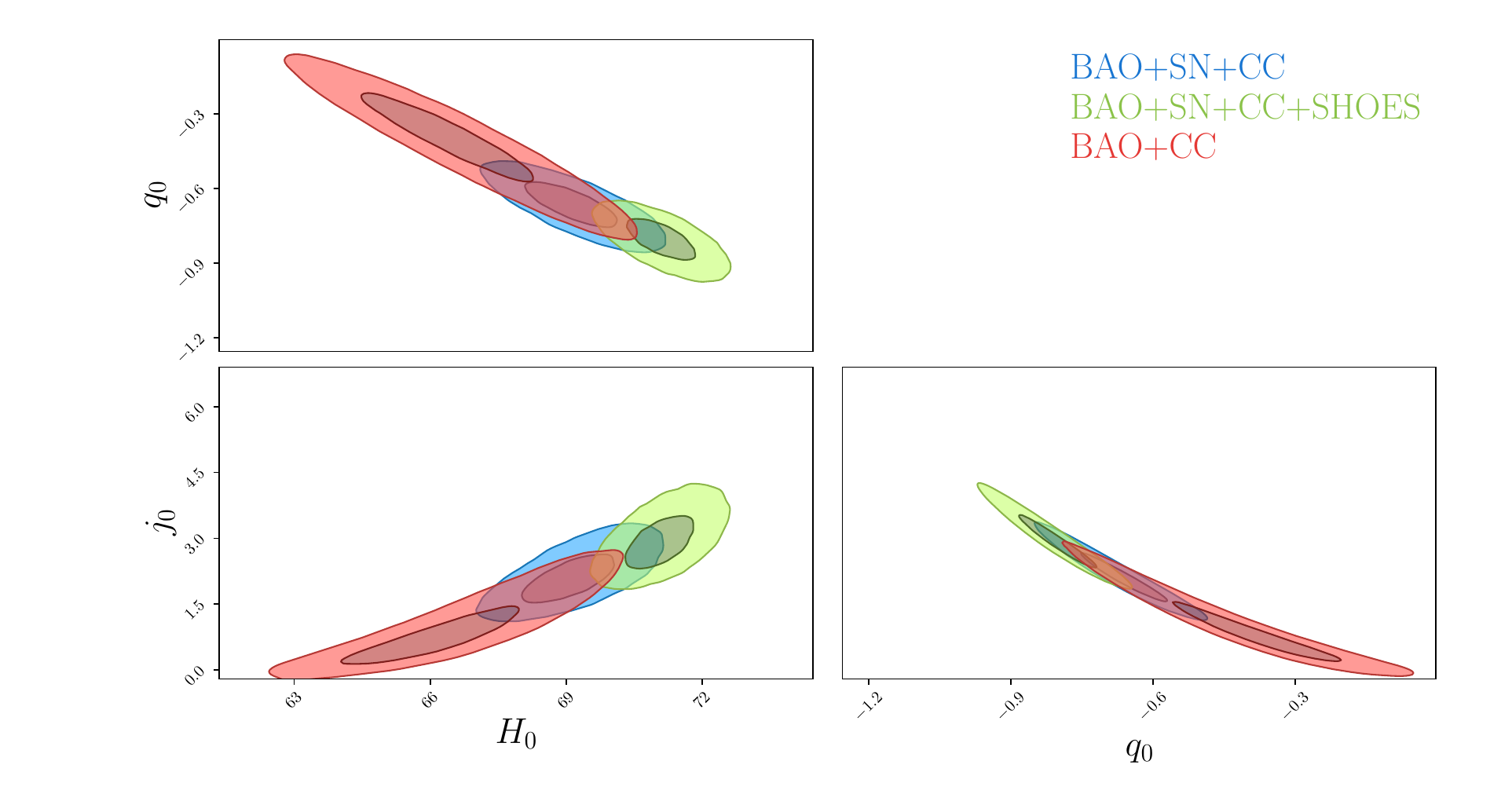} 
\caption{Two-dimensional marginalized confidence regions (68\% and 95\% C.L.) and one-dimensional posterior distribution for the cosmographic coefficients obtained from the BAO\,+\,SN\,+\,CC and BAO\,+\,SN\,+\,CC\,+\,SH0ES joint analyses for the $y$-redshift (2,1) Pad\'e parametrization. The $H_0$ values are expressed in units of km/s/Mpc.}
\label{fig:base2}
\end{center}
\end{figure*} 

In Fig.~\ref{fig:base1}, we show the 68\% C.L. and 95\% C.L. contour regions for the cosmographic series obtained from the joint BAO\,+\,SN\,+\,CC and BAO\,+\,SN\,+\,CC\,+\,SH0ES  analyses. It is important to emphasize that adding the SH0ES prior produces an increase of $H_0$, which, in turn, affects the other cosmographic parameters.

Now, we shall discuss the main results related to the (2,1) Pad\'e parametrization, summarized in Table \ref{tab:mmodel2}. With regard to accuracy  and observational limits on the parameters $H_0$ and $q_0$, we do not notice any significant differences with respect to the $y$-redshift case. In short, we find 1.4\% and 1\% accuracy constraints on $H_0$ from the joint BAO\,+\,SN\,+\,CC and BAO\,+\,SN\,+\,CC\,+\,SH0ES  analyses, respectively. These measurements show, respectively, 2$\sigma$ and 1.4$\sigma$ tensions with the estimates of $H_0$ provided by the SH0ES team. From the perspectives of both Pad\'e and $y$-redshift parametrizations, the constraints on $H_0$ are in accordance with BAO, BBN and CMB obtained in the standard cosmological context \cite{Schoneberg:2022ggi,Philcox:2020vvt,Chen:2021wdi,Nunes:2020hzy}. 

\begin{figure*}
\begin{center}
\includegraphics[width=3.35in]{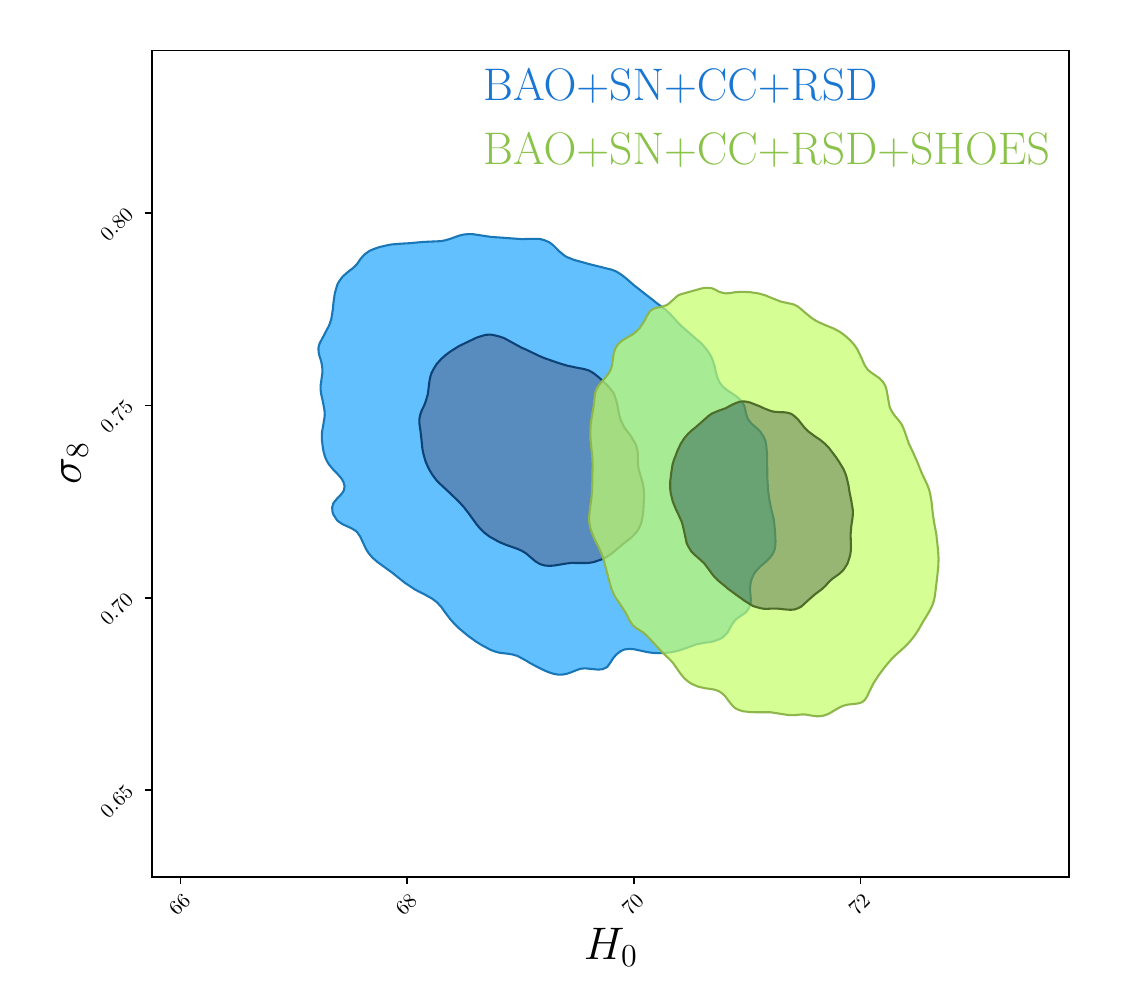} \,\,\,\,\,\,
\includegraphics[width=3.35in]{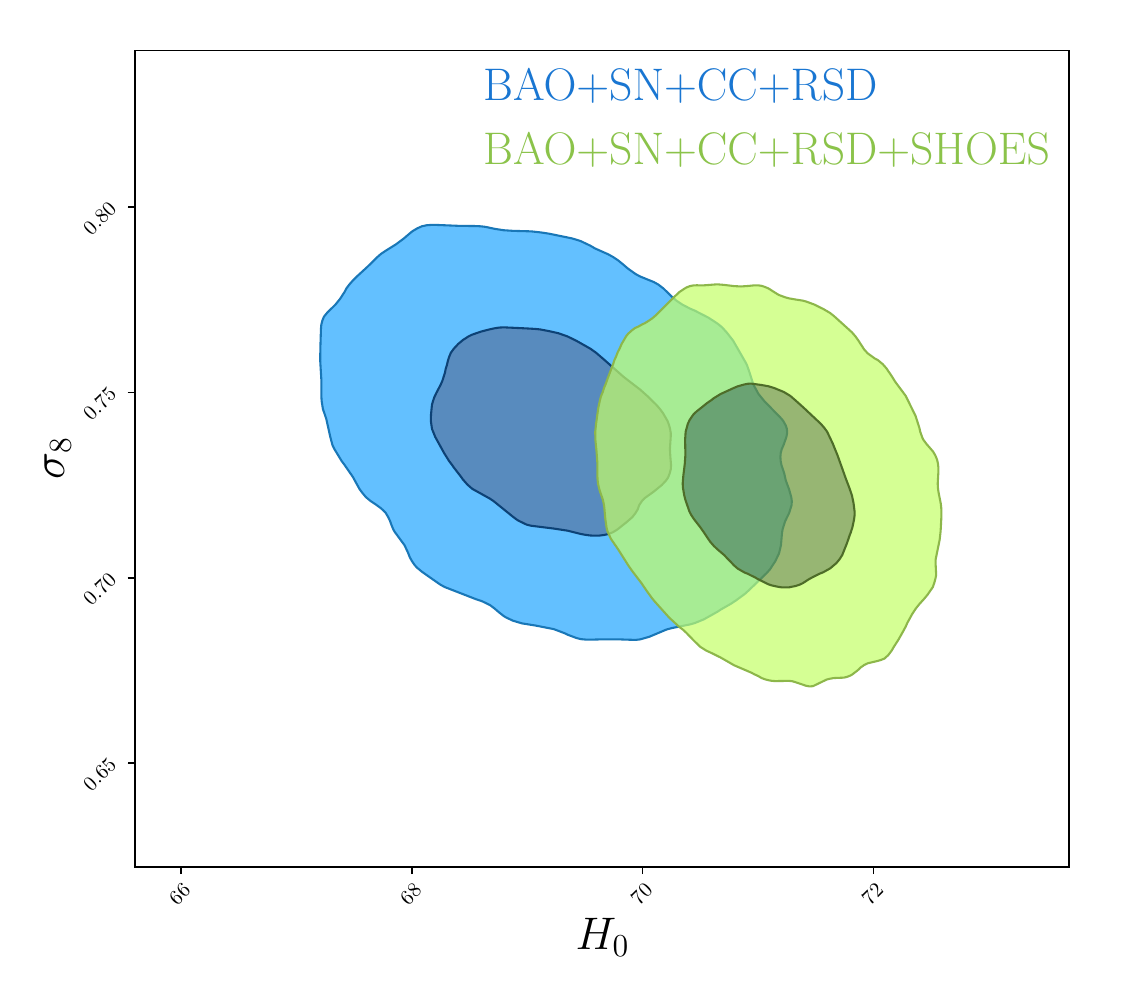}
\caption{68\% and 95\% C.L. contours in the $\sigma_8-H_0$ parameter space, obtained from the dataset combinations BAO\,+\,SN\,+\,CC\,+\,RSD and BAO\,+\,SN\,+\,CC\,+RSD\,+\,SH0ES, for the $y$-redshift (\emph{left panel}) and (2,1) Pad\'e parametrizations (\emph{right panel}).}
\label{fig:s8_h0}
\end{center}
\end{figure*}

Fig.~\ref{fig:base2} shows the 68\% and 95\% C.L. contour regions resulting from for three joint analyses, namely, BAO\,+\,CC, BAO\,+\,SN\,+\,CC and BAO\,+\,SN\,+\,CC\,+\,SH0ES. In particular, the latter provides $j_0$ values that are more than 3$\sigma$ away from the $\Lambda$CDM prediction, i.e., $j_0=1$. However, from the BAO\,+\,CC analysis, we find $j_0=0.59^{+0.93}_{-0.53}$, which is compatible with $\Lambda$CDM at $1\sigma$. Thus, under the $(2,1)$ Pad\'e approximation, when considering the SN sample, the constraints on $j_0$ are not compatible with the standard model. Our result is consistent with the findings of \cite{RodriguesFilho:2017pjx} who, starting from another methodology and dataset, conclude that SN and $H(z)$ data are incompatible with the $\Lambda$CDM model at 2$\sigma$ C.L., and also with each other. Furthermore, using Gaussian process, \cite{Mehrabi:2021cob} find that the jerk parameter is more than 3$\sigma$ away from that of $\Lambda$-cosmology. Therefore, it seems that, from a model-independent point of view, $\Lambda$CDM may be incompatible with the jerk parameter in light of SN data. Similar conclusions are reported by \cite{Luongo:2020aqw,Capozziello:2020ctn}.   

As a guideline, we can compare our findings with the predictions of the flat $\Lambda$CDM model. Specifically, assuming the concordance value $\Omega_{m0}=0.3$, one obtains the following values for the cosmographic parameters: ($q_0$, $j_0$, $s_0$, $l_0$) = ($-0.55$, 1, $-0.35$, 3.115)\footnote{It is worth noting that $j_0=1$ is a theoretical prediction, being independent of any cosmological parameter (see \Cref{appendix}).}. Thus, from the perspective of the $y$-redshift parametrization, our constraints on $j_0$ are compatible with the $\Lambda$CDM predictions. The high-order cosmographic parameters, namely $s_0$ and $l_0$, although poorly constrained, are also fully compatible with the $\Lambda$CDM cosmology.

\begin{figure*}
\begin{center}
\includegraphics[width=3.2in]{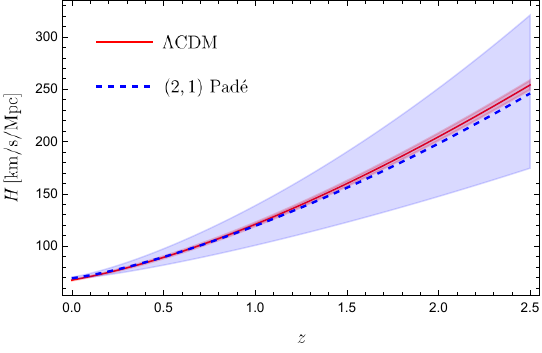} \qquad 
\includegraphics[width=3.28in]{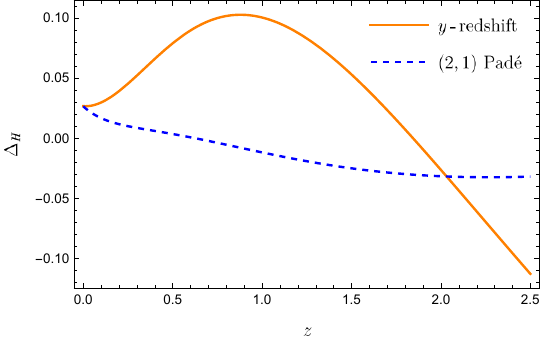} 
\caption{Comparison between the results of our BAO\,+\,SN\,+\,CC\,+RSD joint analysis of the cosmographic parametrizations and those of the $\Lambda$CDM model from Planck \cite{Planck:2018vyg}. 
Left panel: $1\sigma$ reconstruction of the Hubble rate for the (2,1) Pad\'e parametrization and the $\Lambda$CDM model.
Right panel: relative difference of the Hubble rate from the best-fit values of  the $y$-redshift and (2,1) Pad\'e parametrizations, with respect to the best-fit values predicted by the $\Lambda$CDM model.}
\label{fig:H_behaviour}
\end{center}
\end{figure*}

On the left panel of Fig.~\ref{fig:H_behaviour}, we show the statistical reconstruction at the 1$\sigma$ C.L. of the Hubble parameter for the (2,1) Pad\'e model from the BAO\,+\,SN\,+\,CC\,+\,RSD analysis. The latter has been considered due to the fact that it is free of the SHOES $H_0$ prior. Then, we can compare our results with the $\Lambda$CDM predictions inferred from some independent observations. In particular, we anchor the $\Lambda$CDM dynamics to the 1$\sigma$ values obtained from the Planck-CMB data \cite{Planck:2018vyg}. From a statistical point of view, the Pad\'e reconstruction includes the $\Lambda$CDM model as a subset. The error bars for the Pad\'e parametrization appear larger due to the presence of extra free degrees of freedom that are not robustly constrained when compared to a model-dependent analysis based on the $\Lambda$CDM paradigm. Anyway, the cosmic expansion rate given by the Pad\'e parametrization is compatible with $\Lambda$CDM for the entire  redshift range considered here. 
On the other hand, we do not show the $1\sigma$ reconstruction for the $y$-redshift model as this is very degenerate. Such behavior is indeed expected as the high-order cosmographic terms are not properly constrained due to less accurate and few measurements at high redshifts \cite{Capozziello:2019cav,Capozziello:2020ctn}.

Moreover, it is interesting to estimate the relative deviations with respect to the $\Lambda$CDM model. This can be done through the quantity
\begin{equation}
\label{Delta_H}
\Delta_H \equiv \frac{H_i}{H_{\Lambda\text{CDM}}}-1\,,
\end{equation}
where the index $i$ labels the $y$-redshift and the (2,1) Pad\'e parametrizations. 
In this regard, on the right panel of Fig.~\ref{fig:H_behaviour}, we show the relative difference of the Hubble rate for both cosmographic approaches 
with respect to the Planck-$\Lambda$CDM baseline \cite{Planck:2018vyg}. In this case, we assume the best-fit values from the BAO\,+\,CC\,+\,SN\,+\,RSD  analysis, while the lerk parameter is fixed to the reference value expected in the $\Lambda$CDM context, to overcome the unconstrained result of the $y$-redshift parametrization. For the (2,1) Pad\'e parametrization, we find a discrepancy of $\lesssim3\%$ with respect to $\Lambda$CDM over the whole redshift interval under consideration. This discrepancy increases up to $\sim 10\%$ in the case of the $y$-redshift parametrization.

It is possible to quantify the level of tension between two estimates $i$ and $j$ of $H_0$ by means of the simple 1D tension metric, which can be constructed as
\begin{equation}
T_{H_0} \equiv \frac{|H_{0,i}-H_{0,j}|}{\sqrt{\sigma^2_{H_{0,i}}+\sigma^2_{H_{0,j}}}}\,,
\label{eq:1D_estimator}
\end{equation}
measured in equivalent Gaussian standard deviations.
In particular, we find that the results from the BAO\,+\,SN\,+\,CC and BAO\,+\,SN\,+\,CC\,+\,SH0ES combinations are in 1.9$\sigma$ and 1.3$\sigma$ tensions, respectively, with the local measurement by SH0ES team.

Let us now turn our attention to the implications of the cosmographic frameworks under consideration on the parameter $\sigma_8$. In particular, the inclusion of the RSD sample in our  dataset combination allows us to obtain direct constraints on $\sigma_8$. In the context of the $y$-redshift parametrization, from the joint BAO\,+\,SN\,+\,CC\,+\,RSD and BAO\,+\,SN\,+\,CC\,+\,RSD\,+\,SH0ES analyses, we find $\sigma_8=0.737^{+0.029}_{-0.027}$ and $\sigma_8=0.725^{+0.026}_{-0.029}$, respectively (c.f. Table \ref{tab:mmodel1}). Applying the 1D tension metric \eqref{eq:1D_estimator} to the $\sigma_8$ parameter, we note that these estimates are, respectively, at 2.6$\sigma$ and 3$\sigma$ tensions with Planck-CMB data inferred from the $\Lambda$CDM cosmology \cite{Planck:2018vyg}. Our constraints are, however, in agreement with those from weak lensing surveys DES \cite{DES:2021wwk}, KiDS \cite{KiDS:2020suj,Amon:2022ycy} and galaxy clustering measurements \cite{Kobayashi:2021oud,Yuan:2022jqf}. 

Very similar results are found when using the (2,1) Pad\'e parametrization. We can see this in Fig.~\ref{fig:s8_h0}, which shows the $\sigma_8$-$H_0$  parameter space, together with the corresponding 68\% C.L. and 95\% C.L. contours, obtained from the combinations BAO\,+\,SN\,+\,CC\,+\,RSD and BAO\,+\,SN\,+\,CC\,+\,RSD\,+\,SH0ES for both cosmographic approaches. Therefore, we can conclude that the amplitude of matter fluctuations analyzed through cosmographic expansion at late times shows a significant tension with CMB measurements. 
Finally, we note that the effects of RSD data provide an improvement in the accuracy of the $j_0$ parameter, whose constraints are, in any case, much more stringent for the (2,1) Pad\'e approximation compared to the $y$-redshift parametrization.

\section{Final remarks}
\label{sec:conclusions}

In this work, we showed how model-independent approaches may reveal a promising tool for investigating tensions among cosmological parameters when inferred from different datasets. 
For our purposes, we specifically considered the cosmographic technique, based on a series expansion of a cosmological observable around $z = 0$, and describing the late-time evolution of the universe through a set of kinematic parameters to be constrained directly by data. 

In particular, motivated by suitable properties able to heal the convergence problems typical of standard cosmography, we considered the improved $y$-redshift and Pad\'e parametrizations of the Hubble expansion rate to accurately describe the dynamics of the universe independently of cosmological model choices. Under these theoretical frameworks, we chose a robust data sample  consisting of recent BAO, SN, CC and RSD measurements in the redshift range $z \in (0, 2.3)$ to explore possible inconsistencies with the $\Lambda$CDM model predictions.

We thus performed an MCMC numerical analysis on the combination BAO\,+\,SN\,+\,CC\,+\,RSD, obtaining $H_0$ and $\sigma_8$ estimates with an accuracy of $\sim$1.4\% and $\sim$3.7\%, respectively. It is worth stressing that our model-independent constraints are competitive with those inferred by assuming specific cosmological backgrounds.
Without including the SH0ES prior on $H_0$, our measurements are at 2$\sigma$ tension with the
local measurements, which represents a significant reduction compared to the Planck-CMB estimate assuming the $\Lambda$CDM model. 
On the other hand, our measurements on $\sigma_8$ are at 2.6$\sigma$ tension with the Planck-$\Lambda$CDM cosmology. 

Furthermore, we found that the jerk parameter can deviate $>3\sigma$ from the prediction of the $\Lambda$CDM model. Since different dark energy models predict different values and behaviors for $j_0$, a confirmation of our results by future data, possibly through further improved cosmographic modeling, may pose a new challenge and internal tension within the standard cosmological model. 
Also, since heterogeneous measurements have been combined to constrain $H_0$ and $\sigma_8$, it might be worth analyzing the impact of different systematic errors and statistical weights on the final results. We leave this subject for future work.

\acknowledgements
The authors would like to thank Salvatore Capozziello, Orlando Luongo and Eoin Colgain for useful discussions on the topic of cosmography. R.D. acknowledges the financial support of INFN - Sezione di Napoli,  \emph{iniziativa specifica} QGSKY. R.C.N thanks the CNPq for partial financial support under project No. 304306/2022-3. The authors also thank the anonymous referee for his/her valuable comments and suggestions.

\appendix
\section{$\Lambda$CDM prediction for the cosmographic parameters}
\label{appendix}
In this appendix, we show the theoretical expressions of the cosmographic parameters as expected for the flat $\Lambda$CDM model, whose Hubble expansion rate at late times is given by
\begin{equation}
    H(z)=H_0\sqrt{\Omega_{m0}(1+z)^3+1-\Omega_{m0}}\,.
    \label{eq:HLCDM}
\end{equation}

To do that, we start from the definitions \ref{eq:CS} and convert the time derivatives into derivatives with respect to the redshift by means of the relation
\begin{equation}
    dt=-\frac{dz}{(1+z)H(z)}\,.
\end{equation}
In so doing, one finds the following expressions:
\begin{align}
    q(z)&= -1+(1+z)\dfrac{H'}{H}\,, 
    \label{eq:q(z)} 
\\
    j(z)&= 1-2(1+z)\frac{H'}{H}+(1+z)^2\left(\frac{H'}{H}\right)^2+(1+z)^2\frac{H''}{H}\,, 
    \label{eq:j(z)}
    \end{align}
\begin{align}
    s(z)&= 1-3(1+z)\dfrac{H'}{H}+3(1+z)^2 \left(\frac{H'}{H}\right)^2-(1+z)^3 \left(\frac{H'}{H}\right)^3 \nonumber \\
    &+(1+z)^2\frac{H''}{H}-(1+z)^3\frac{H^{(3)}}{H}-4(1+z)^3\frac{H' H''}{H^2}\,,
    \label{eq:s(z)}
\\
    l(z)&= 1-4(1+z)\dfrac{H'}{H}+6(1+z)^2\left(\frac{H'}{H}\right)^2-4(1+z)^3\left(\frac{H'}{H}\right)^3 \nonumber \\
    &+(1+z)^4\left(\frac{H'}{H}\right)^4+2(1+z)^2\frac{H''}{H}+(1+z)^3\frac{H^{(3)}}{H} \nonumber \\
    &+(1+z)^4\frac{H^{(4)}}{H}-(1+z)^3\frac{H'H''}{H^2}+7(1+z)^4\frac{H'H^{(3)}}{H^2} \nonumber  \\
    &+11(1+z)^4 \frac{{H'}^2 H''}{H^3}+4(1+z)^4\left(\frac{H''}{H}\right)^2.
    \label{eq:l(z)}
\end{align}

Thus, inserting \Cref{eq:HLCDM} into \Cref{eq:q(z),eq:j(z),eq:s(z),eq:l(z)}, and evaluating the results at $z=0$, we obtain
\begin{align}
    q_0& =-1+\dfrac{3}{2}\Omega_{m0}\,, \\
    j_0& =1\,, \\
    s_0&=1-\frac{9}{2}\Omega_{m0}\,, \\
    l_0&= 1+3\Omega_{m0}+\dfrac{27}{2}\Omega_{m0}^2 \,.
\end{align}
Notice that all the cosmographic parameters are independent of the value of the Hubble constant. Moreover, $j_0$ is fixed to the unity in the $\Lambda$CDM scenario, regardless of the $\Omega_{m0}$ value.

\bibliography{references}

\end{document}